\documentstyle[aps,prb,multicol,epsf]{revtex}
\def\Tr{\mbox{Tr}}
\def\D{\mbox{D}}

\def\Im{\mbox{Im}}
\def\max{\mbox{max}}
\def\I{\mbox{I}}
\def\M{\mbox{M}}
\def\S{\mbox{S}}

\def\narrow{\begin{multicols}{2} \global\columnwidth20.5pc}
\def\wide{\end{multicols} \global\columnwidth42.5pc}
\def\top#1{\vskip #1\begin{picture}(290,80)(80,500)\thinlines \put(
65,500){\line( 1, 0){255}}\put(320,500){\line( 0, 1){5}}\end{picture}}
\def\bottom#1{\vskip #1\begin{picture}(290,80)(80,500)\thinlines \put(
330,500){\line( 1, 0){255}}\put(330,500){\line( 0, -1){
5}}\end{picture}}

\title {
Shot Noise in Disordered Junctions:
Interaction Corrections.
}
\author{D.B. Gutman and Yuval Gefen }
\address{
Department of Condensed Matter Physics,
The Weizmann Institute of Science,
\newline
76100 Rehovot, Israel}
\begin{document}
\date{\today}
\maketitle
\begin{abstract}
We study current correlation functions in a diffusive junction out of equilibrium.
We calculate corrections to the electric current and to the
zero frequency shot noise due to electron-electron interactions.
Contrary to the equilibrium situation
(where the corrections to the current and to the current noise are related through the fluctuation-dissipation theorem (FDT)),
these two quantities behave differently:
the correction to the electron current
are governed by the largest of temperature and applied voltage, while
the correction to the shot noise is always governed by the temperature.
\end{abstract}

\begin{multicols}{2}
\section{Introduction}
\label{sec_1}
The physics of  non-equilibrium mesoscopic systems was the  subject of an extensive research for the last decade \cite{Kogan_book,Blanter,Beenaker_proc}.
Khlus \cite{Khlus}, who employing a semiclassical approximation
considered  a multichannel two-terminal junction, was the
first  to notice that the non-equilibrium current
noise in such a system strongly depends on the transparency
of the barrier, and vanishes in the limit of completely transparent junction.
 This is in contrast to the conductance which remains finite in this limit.
Later  Lesovik \cite{Lesovik} has shown that for coherent transport through
a multi-channel two-terminal system the zero frequency current noise
\cite{zero_freq}
is given by
\begin{eqnarray}&&
\label{Les}
\S(0)\!=\!\frac{e^2}{2 \pi \hbar}\int dE\sum_n\{T_n(E)[f_L(E)(1\!-\!f_R(E)]\!+ \nonumber \\&&
\!f_R(E)[1\!-\!f_L(E)]\!-\!T_n^2(E)[f_L(E)\!-\!f_R(E)]^2\} \, \, .
\end{eqnarray}
This was derived using the Landauer \cite{Landauer} scattering states approach.
Here  $f$ is the grand-canonical Fermi-Dirac distribution function,
$L$ and $R$ denote the left and right reservoirs respectively,
taken at different values of the chemical potential, and
$\{T_n\}$ are the channel transparencies.
This result was later generalized by B\"uttiker\cite{Buttiker} to include
the multi-channel multi-terminal case  {and experimentally confirmed}
by Reznikov {\it et. al.} \cite{Reznikov1}.
The zero temperature limit of eq.(\ref{Les}) yields the
shot noise:
\begin{eqnarray}
\label{equation2}
\S(0)=\frac{e^2}{2\pi\hbar}\sum_n T_n(1-T_n)eV \,\,.
\end{eqnarray}
The remarkable dependence of the shot noise on the quasiparticle
charge ( as was first emphasized by Kane and Fisher \cite{Kane})
allowed the first direct measurement of fractional charge in the
context of FQHE by Reznikov {\it et. al.}\cite{Resnikov2} and by
Saminadaya {\it et. al.} \cite{Glattli} The suppression of the
shot noise for open channels ( in full accordance with  the
semiclassical result \cite{Khlus}, eq. (\ref{equation2})) is a
manifestation of correlations among the electrons, arising from
Pauli
 exclusion principle.
In disordered systems one usually cannot control (or obtain
information about) the individual channels; instead it is useful
to consider average quantities. For coherent diffusive systems
(smaller than the inelastic, dephasing and the localization
lengths) the various moments of the current-current correlation
function have been calculated employing random matrix theory (RMT)
\cite{Beenakker}. Interestingly enough, it turned out that the
fermionic suppression of the noise present in ballistic systems is
manifest in the configuration averaged noise (in diffusive
systems) as well: the shot noise is suppressed by a factor of
$1/3$ as compared with the result expected for classical
particles.

Other than the Landauer scattering states approach, one may use
the kinetic equation, or the so-called Kogan-Shulman
approach\cite{Shulman}. It is valid for systems where the dynamics
(but not necessarily the statistics) of the particles is
classical. Comparing with the scattering states approach, the
latter is not restricted solely to non-interacting electrons.
Unfortunately, it is limited by the usual conditions of
applicability of the kinetic equation: it does not yield quantum
interference effects, and  cannot be used for calculation of the
cummulants higher than two. To overcome the latter shortcomings
one may employ the Keldysh formalism \cite{Altshuler-Levitov}.

In the present analysis we find it convenient to employ this
formalism in the form of a non-linear  sigma-model. This machinery
was found  to be an effective tool for describing the low energy
physics of  disordered electronic systems\cite{Larkin}. We will
not attempt to review here the various technical steps involved in
the derivation of the model, but rather recall some of the  main
ingredients involved. So long  as non-interacting electrons are
concerned, the super-symmetric sigma-model\cite{Efetov} turns out
to be a particularly  suitable tool for calculation. The inclusion of
electron-electron interaction appears, though, to render this
approach  infeasible. Interaction among the electrons could be
included within the  the replica sigma-model, which has been
introduced by Finkelstein for the study  of
disordered Fermi liquids \cite{Fin}. As has recently been
demonstrated  by Kamenev and Andreev\cite{Kamenev} and
independently by Chamon, Ludwig and Nayak \cite{Chamon}, there is
an alternative approach of constructing a sigma-model for
interacting electrons - the so-called Keldysh sigma-model.

To wrap up this introduction, we find it useful to  define some of
the physical regimes to be discussed below. We first stress that
the notion of high/low frequency carries two meanings. As far as
the diffusive motion of the electrons (viewed as independent
particles) is considered, one can refer to the frequency as
''low'' if it is smaller than the Thouless energy of the system,
$E_{Th}=\hbar D/L^2$. As we study   the case of interacting
electrons, another criterion arises, relating to the propagation
of electro-magnetic fields in the system. Consider the simplest
situation when the effects of external screening (by, e.g.,
external gates) can be neglected. It is readily seen that the
solution of the Maxwell equations (or the self-consistent solution
of the Maxwell equations combined with the kinetic equations, cf.
Section \ref{sec_3} below) depends on the effective dimensionality
of the sample. For a standard three dimensional geometry the
frequency at which the current noise is measured ($\omega$) should
be compared with the inverse Maxwell time ($\omega_M=4\pi\sigma$).
At $(\omega \ll \omega_M)$ the ''cross-coordinate'' current
correlation (measured at different cross-sections) is independent
of the spatial coordinates (and their separation, $l_{x,x'}$).
Having this electrodynamics in mind, it is this limit which should
be referred to as  ``low frequency`` . For quasi-two dimensional
films  this regime is realized under the condition:
 $ \omega \ll \kappa_2 D/l_{x,x'}$ where the inverse screening length
$\kappa_2= 2\pi e^2 \nu $. The effects of external screening may
modify the effective electron interaction,
\cite{Nagaev3,Nagaev2,Likharev}. Our point, though, is  that quite
generally  there is some frequency, below which the correlation
function does not depend on the spatial coordinates, i.e. it is a
constant as  function of $|l_{x,x'}|$.   In the
case of interacting electrons, in addition to the long wave-length
propagation of the electro magnetic field, one also needs to
account for inelastic collisions among the electrons. We thus
introduce another dimensionless parameter, namely  the ratio
between the system's length, $L$, and the inelastic length
$l_{in}$.

The outline of the present work is the following: In the next
section  we consider non-interacting spinless electrons in a
weakly disordered conductor neglecting weak (and in two-dimensions
strong as well) localization effects. The latter can be justified
once the quantum coherence of the electrons is destroyed by
certain dephasing mechanisms (such as fluctuating electro-magnetic
fields), or is restricted by the finite size of the system. In
other words,  we consider a situation where the localization
length, $\xi$, exceeds either the dephasing length $l_\phi$ or the
system size ($\xi \gg \max\{l_\phi, L\}$). We extend the Keldysh
sigma-model formalism to the case  of systems out of equilibrium
(including the effect of restricted geometries). In Section
\ref{sec_3} we review the calculation of the current noise (in and
out of equilibrium), discussing both the low and the high
frequency limits, paying special attention to the non-homogeneity
(in space) of the current fluctuation in the later case. In
Section \ref{sec_4}  we consider the effect of electron-electron
interactions. We review results obtained from the kinetic equation
approach, referring briefly to the limits of $l_{in} \ll L$ and
$l_{in} \gg L$. We then focus on the latter limit  presenting
results for the effect of electron-electron interactions on the
noise which go beyond the kinetic equation (eqs. \ref{f2} and
\ref{cor1}). Our results are analogous to the Altshuler-Aronov
corrections \cite{AA,AAL} found earlier for equilibrium noise.
This is done using a model short-ranged interaction. The general
expressions obtained for the noise are then analyzed in the
interesting case of two-dimensional geometries. We show that
electron-electron interaction leads to the suppression of the
non-equilibrium current noise which is stronger than the
suppression of the current itself - a manifestation of many-body
correlation effects. Eqs. (\ref{e29}, \ref{r2}   and  \ref{f8})
are the main results of our analysis. The three appendices include
some technical parts of the derivation. The analysis presented
here is an extended account of an earlier work\cite{GG}.
\section{Keldysh Formalism: non-interacting electrons in diffusive system}
\label{sec_2}
The theory for quantum systems out of equilibrium was
suggested simultaneously by Kadanoff and Baym and by  Keldysh \cite{Keldysh}.
This technique has been widely used on the level of  diagrammatic calculations \cite{Lifshitz} .
Such techniques can also be developed for the study of weakly disordered
samples \cite{Rammer}.
Instead of using here a diagrammatic technique,
we will present a path-integral description of the problem.
For free fermions the action can be written as
\begin{eqnarray}
\label{ferm_act} S=\int d{\bf r} \int_c dt {\cal L} \,\, ,
\end{eqnarray}
where the Lagrangian density is given by
\begin{eqnarray}
{\cal L}=\overline\Psi[\hat{G}_0^{-1}-U_{dis}]\Psi \,\, .
\end{eqnarray}
Here the Schr\"odinger operator of an electron of mass $m$ in the presence of
a vector potential $e{\bf \tilde a}/c$ is given by
\begin{eqnarray}
\hat{G}^{-1}_0=i\hbar\frac{\partial}{\partial t}+\frac{\hbar^2 }{2 m}\left(\nabla-{\bf \tilde a}\right)^2+\mu\,\, .
\end{eqnarray}
The energy is measured from the chemical potential $\mu$.
The action now contains an integral over space and an integral in time
over a  Keldysh contour~(see~Fig~\ref{fig7}).
In analogy with ordinary diagrammatic techniques  one defines an
electron Green's function, i.e. a time ordered correlation
function
\begin{eqnarray}
\label{e13}
\tilde{G}_{\alpha\beta}(r,t;r',t')=-i
\langle T_c\Psi(r,t_\alpha) \Psi^\dagger(r',t'_\beta)\rangle
\end{eqnarray}

Here $T_c$ is a time ordering operator along a Keldysh contour,
the indices $\alpha, \beta$ denote the branches of the Keldysh
contour (minus or plus for the upper and lower branch of the
contour respectively). Here $\langle \, \rangle$ denotes averaging
with respect to the action eq. (\ref{ferm_act}).

\begin{minipage}{3.1in}
\begin{figure}
\epsfysize=.5 in
{\makebox(0,-5){${-\infty}$}}
{\makebox(0, -45){${\infty}$}}
\centerline{\epsfbox{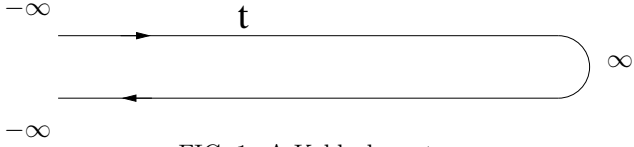}}
{\makebox(0,0){${-\infty}$}}
\centerline{\caption{A Keldysh contour}}
\label{fig7}
\end{figure}
\end{minipage}

It is also useful
to introduce the generating functional
\begin{eqnarray}
\label{aa3}
Z[{\bf \tilde a}]=\int {\cal D}\Psi \exp(iS[\Psi,{\bf\tilde a}]/\hbar) \, .
\end{eqnarray}
One notices that the generating functional (\ref{aa3}) is
identically equal to unity, so long as the external source is the
same on the upper and lower branches \cite{cancelations}
($\tilde{a}_-(t)=\tilde{a}_+(t)$). For this reason one may
explicitly perform the averaging of the generating functional over
disorder. Below we consider short-range and  $\delta$-correlated
disorder, which leads to following definition of disorder
averaging:
\begin{eqnarray}
\langle  \, \rangle =\int {\cal D} U_{dis} \exp\left(-\pi \nu \tau
\int dr\,U_{dis}^2(r) \right) \,\, ,
\end{eqnarray}
where $\tau$ and $\nu$ are the elastic mean free time and  the single-particle density of states respectively.
When  applied to the generating functional it produces an effective non-linear action
\begin{eqnarray}&&
\label{aa10}
S=\int d{\bf r}\int_{-\infty}^{\infty} dt \overline\Psi \hat{G}_0^{-1} \sigma^{(3)}\Psi-\nonumber \\&&
(4 \pi \nu)^{-1} \int dr \bigg[\int_{-\infty}^{\infty} dt \overline\Psi (r,t)\sigma^{(3)} \Psi(r,t)\bigg]^2
\,\, ,
\end{eqnarray}
where
\begin{eqnarray}
\sigma^{(0)}=\left(\matrix {1 & 0\cr 0 & 1 \cr}\right),  \,
\sigma^{(3)}=\left(\matrix {1 & 0\cr 0 & -1 \cr}\right).
\end{eqnarray}
The non-linear term in eq.(\ref{aa10}) can be decoupled with an
auxiliary bosonic (a  $2\times2$  matrix )
field $\tilde{Q}(r,t,t')\equiv\tilde{Q}_{r,t,t'}$,
\end{multicols}
\top{-3cm}
\parbox{6in}{
\begin{eqnarray}&&
\exp \left\{\!-(4\pi\hbar \nu\tau)^{-1}\!\! \int \!\! d{\bf r}
\left[\int\!dt\overline\Psi({\bf r},t)\sigma^{(3)}\Psi({\bf
r},t)\right]^2\right\}
=\nonumber \\ &&
\!\int{\cal D}\tilde{ Q}
\exp\left\{\!-\int d{\bf r}dtdt'\left[ \frac{\pi\nu\hbar}{4\tau}
\Tr\,\tilde{ Q}_{tt'}({\bf r})\tilde{ Q}_{t't}({\bf r})
+\frac{1}{2\tau} \overline\Psi({\bf r},t)\tilde{Q}_{tt'}({\bf
r})\sigma^{(3)} \Psi({\bf r},t') \right]\right\} .
\end{eqnarray}}
\begin{multicols}{2}
As a result of this Hubbard-Stratonovich transformation one obtains
the following action:
\begin{equation}
\label{g1}
iS[\tilde{Q}]=-\frac{\pi\hbar \nu}{4\tau} \Tr \{ \tilde{Q}^2\} + \Tr \ln[\left(\hat{G}^{-1}_0 +\frac{i}{2\tau}\tilde{Q}\right)\sigma^{(3)}]
\,\, ,
\end{equation}
It follows from equation (\ref{e13}) that
the components of the matrix Green function are linearly related
\begin{eqnarray}
\tilde{G}_{--}+\tilde{G}_{++}=\tilde{G}_{-+}+\tilde{G}_{+-} \, .
\end{eqnarray}
It is thus convenient to perform a rotation in the Keldysh space
\begin{eqnarray}
\label{rot1}
G=R\sigma^{(3)}\tilde{G}R^\dagger \,\,\, ,\,\, {\bf a}=\frac{1}{\sqrt 2}R^\dagger{\bf  \tilde a} \,\, .
\end{eqnarray}
Here the unitary rotation matrix $R$ is given by
\begin{equation}
\label{aa4}
R=\frac{1}{\sqrt{2}}\left(\sigma^{(0)}-i\sigma^{(2)}\right)=\frac{1}{\sqrt{2}}
\left( \begin{array}{rr}
1&-1 \\ 1&1 \end{array} \right)\, ,
\end{equation}
yielding (cf. eq. (\ref{g1}))
\begin{eqnarray}
\label{g2}
iS[Q]=-\frac{\pi\hbar\nu}{4\tau}\Tr \{ Q^2\} + \Tr\ln[\hat{G}^{-1}_0 +\frac{i}{2\tau}Q]
\,\, .
\end{eqnarray}
Clearly, the resulting action is exactly equivalent to the original
one, eq. (\ref{aa10}).
However,  it turns out that in many physical problems the fast variations of
the matrix $Q$ in space and time are unimportant, and we may integrate them out.
In the weak disorder case
$(\epsilon_F\tau \gg \hbar$, where $\epsilon_F$ is a Fermi energy)
the main contribution comes from the minimum of the action,
justifying the use of a saddle-point approximation.
Varying the action with respect to $Q$,  one arrives at the following saddle-point equation
\begin{eqnarray}
\label{e22}
\pi\hbar\nu Q =\frac{i}{\hat{G}^{-1}_0\sigma^{(0)}+\frac{i }{2\tau}Q}
\,\, .
\end{eqnarray}
Taking the  Fourier transform with respect to $t-t'$ one can show
that the matrix
\begin{eqnarray}
\label{e23}
\Lambda({\bf r},\epsilon) =\left(\matrix {1& 2F(\epsilon) \cr
0& -1\cr}\right)
\end{eqnarray}
( for an arbitrary $F$) is a solution of equation (\ref{e22}). Apart from
$F(\epsilon)$ which needs to be determined, the
solution for  the saddle-point (eq. \ref{e23}) is not unique.
Indeed, the action (\ref{g2}) is invariant under any  homogeneous  unitary transformation
\begin{eqnarray}
\label{e14}
Q=U^\dagger \Lambda U
\,\, .
\end{eqnarray}

To calculate the fluctuations around the saddle point (\ref{e23})
we will distinguish between longitudinal and transversal
fluctuations, the latter do not modify the value of $Q^2$. The
spectrum of the longitudinal fluctuations has a gap; for this
reason they are irrelevant for the low energy physical properties
of the system (although they do affect the bare values of the
parameters). The gapless transversal fluctuations (soft modes)
should be taken into account.

We now focus on the long wavelength  fluctuations around the
saddle point $\Lambda$. To study such fluctuations we now extend
our discussion and assume the matrix $U$ in  eq. (\ref{e14}) to be
a slow function of the spatial coordinate. One may now expand the
action in gradients of   $U({\bf r})$. We thus obtain a
new action in a form which is usually referred to as
\cite{Fradkin} a non-linear sigma model
\begin{eqnarray}
\label{e6}
iS[Q]=-\frac{\pi\hbar\nu}{4}\Tr[D \left( \nabla Q \right)^2 +4i \hat{\epsilon} Q]\,\,,
\end{eqnarray}
where the integration over the field $Q$ is now restricted by the  non-linear constraint
\begin{eqnarray}
\label{nlc}
Q^2=1
\,\, .
\end{eqnarray}

One may derive an equation of motion ( quantum kinetic equation
\cite{Kamenev}), which is compatible with eq. (\ref{nlc}):
\begin{eqnarray}&&
\label{z10}
D\nabla(Q\nabla Q)+i[\hat{\epsilon},Q]=0 \,\, .
\end{eqnarray}
We now  use the following parameterization:
\begin{eqnarray}&&
\label{e24}
Q=\Lambda\exp\left(W\right) \nonumber\,\, , \\&&
\Lambda W+W \Lambda=0 \,\, ,
\end{eqnarray}
where the matrix $W_{x,\epsilon,\epsilon'}$, in turn, is
parameterized as follows
\begin{eqnarray}
W_{x,\epsilon,\epsilon'}\!=
\!\left(
\matrix {F_{x,\epsilon}
\bar{w}_{x,\epsilon,\epsilon'} & -w_{x,\epsilon,\epsilon'} + F_{x,\epsilon}\bar{w}_{x,\epsilon,\epsilon'}F_{x,\epsilon'}\cr-\bar{w}_{x,\epsilon,\epsilon'}  & - \bar{w}_{x,\epsilon,\epsilon'}F_{x,\epsilon'}\cr}
\right),
\end{eqnarray}
and where the fields $F$, $w$ and $\bar{w}$ are unconstrained.
Note that $\Lambda$ is parameterized by $F$ as well, (cf. eq.(\ref{e23})).
Expanding the action  in  soft mode fluctuations around the
saddle point  we obtain,  to second order in
\cite{pert_theory}   $w, \bar{w}$:
\begin{eqnarray}
iS[W]=iS_0[W]+iS_1[W]+iS_2[W] \,\, .
\end{eqnarray}
The zeroth order term is identically equal to zero
\begin{eqnarray}
iS_0[W]=0 .
\end{eqnarray}
This implies that, in the absence of external sources, the
generating functional is equal to unity even on the level of the
saddle point equation. The term linear  in $W$ is given by
\begin{eqnarray}&&
\label{e56}
iS_1[W]=\pi \nu \hbar \Tr\{
i \hat{\epsilon}(\bar{w}F-F\bar{w})+D\bar{w}\nabla^2 F\}.
\end{eqnarray}
Selecting  $\Lambda$ to be a saddle point of the action
(\ref{e6}), which requires the  part linear in $W$ to vanish as
well,  we obtain
\begin{eqnarray}&&
\label{e67}
i(\epsilon_1-\epsilon_2)F_{x,\epsilon_1,\epsilon_2}-D\nabla^2
F_{x,\epsilon_1,\epsilon_2}=0 \,\, .
\end{eqnarray}
The function $F(x,\epsilon)$ is related to the single particle
distribution function $f(x,\epsilon)$
\begin{eqnarray}&&
\label{f_F} F(x,\epsilon)=1-2f(x,\epsilon) \,\, .
\end{eqnarray}
Indeed, eq.(\ref{e67}) coincides with the Boltzmann equation
within the diffusion approximation for $f$. However, since
eq.(\ref{e67}) is a linear homogeneous equation, $F$ may be
related to $f$ only up to  multiplicative factors. To
establish the relation, eq.(\ref{f_F}), one needs to resort to
the inhomogeneous version of eq (\ref{e67}), where the r.h.s. is
replaced by a collision integral due to electron-electron
interactions. This indeed has been done in Ref.\cite{Kamenev} (
cf. with their eq.(138)).

 We can describe what we have done above
slightly differently. We note that the parameterization,
eq.(\ref{e24}), is a particular choice \cite{Jacob} satisfying the constraint,  eq.( \ref{nlc}). We may look for a
particular case for which $\Lambda$ (which, to begin with, was a
saddle point of the original action, eq. (\ref{g2})) is a saddle
point of the non-linear $\sigma$-model action, eq. (\ref{e6})
(i.e., $S_1[W]=0$). This leads to  equation ( (\ref{e67})). On
the other hand , we know that one can employ the Keldysh technique
perturbatively, and relate $\Lambda$ (with the parameterization of
eq. (\ref{e24})) to $\hat{G}(r,r)$ ( where $\hat{G}$ is the
Green's function in the Keldysh space), with the $(1,2)$ element
thereof related to the distribution function $f$. This justifies
referring to eq.(\ref{e67}) as a kinetic equation.  To consider
larger deviations from the saddle point $\Lambda$ , one needs to
consult the exact saddle point equation (\ref{z10}).

Equation (\ref{z10}) (or \ref{e67}) should be supplemented by
appropriate  boundary conditions at the endpoints of the sample.
The quadratic fluctuations around the saddle point are given by
\begin{eqnarray}&&
\label{e55}
iS_2[W]=\frac{\pi\nu\hbar}{2}
\bigg[\bar{w}_{x,\epsilon,\epsilon'}
[-D\nabla^2+i(\epsilon-\epsilon')]w_{x,\epsilon',\epsilon}
-\nonumber \\&&
D \nabla F_{x,\epsilon}\bar{w}_{x,\epsilon,\epsilon'}\nabla F_{x,\epsilon'}
\bar{w}_{x,\epsilon',\epsilon}
\bigg]\,\,.
\end{eqnarray}
This  action describes the dynamics of the fluctuating fields $w$
and $\bar{w}$. The dynamics  may be  affected by   the last term
on the r.h.s. of eq.(\ref{e55}), accounting for deviations from
thermal equilibrium.

We now specialize to a concrete geometry and specific boundary
conditions. We consider a system made of two clean ( i.e., no disorder)
and stationary
reservoirs, connected to each other through a diffusive bridge as
shown in Fig.(\ref{fig3}).

\vspace {1cm}
\begin{minipage}{3.1in}
\begin{figure}[p]
\epsfysize=.8 in
\centerline{\epsfbox[80 10 230 100]{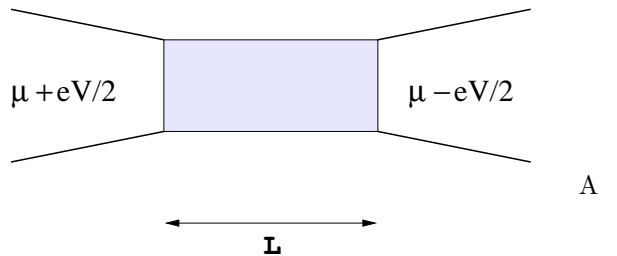}}
\vspace {.5cm}
\caption{A diffusive bridge (the shaded area)  between two reservoirs}
\label{fig3}
\end{figure}
\end{minipage}
\vspace{1cm} A non-zero bias  ($eV \neq 0$) gives rise to a finite
d.c. current  through the bridge. Hereafter we assume that the
typical energy relaxation time is much larger than the
time-of-flight of the electrons through the micro-bridge,
resulting in the following stationary diffusion equation
\begin{eqnarray}
\label{e68}
D \nabla^2 F(r,\epsilon)=0.
\end{eqnarray}
Note that under the conditions specified above, eq. (\ref{e68})
contains no energy collision integral and no time derivative.
Furthermore,  we assume that the typical time  a particle spends
in the reservoir is much larger than the energy relaxation time,
so that the single-electron distribution function is subject to
local equilibrium boundary conditions at the edges, with the
respective chemical potentials as indicated in Fig. (\ref{fig3}).
It turns out that the solution to eq. (\ref{e68}) consistent with
such  boundary conditions is\cite{Nagaev1}
\begin{eqnarray}
\label{dist1}
F(\epsilon,x)\!=\!\left(\frac{x}{L}\right)F_0(\epsilon\!+\!eV/2)\!+\!\left(1\!-\frac{x}{L}\right)F_0(\epsilon-eV/2).
\end{eqnarray}
Here
\begin{eqnarray}
F_0(\epsilon)=\tanh\left(\frac{\epsilon}{2T}\right) \,\,.
\end{eqnarray}

As we have prepared the tools for our analysis of non-equilibrium noise,
we are now set to construct the correlation functions for
$w$ and $\bar{w}$.
It is convenient to define the diffusion propagator as
\begin{eqnarray}
\label{e31}
\left(-iw+ D\nabla^2 \right ) \D(r,r', \omega)=\frac{1}{\pi \nu\hbar}\delta(r-r') \,\, ,
\end{eqnarray}
with the boundary conditions defined on the edges
\cite{AA,diffusion_kineq}. Since no current may flow through
infinite walls , the derivative of the diffusion propagator must
vanish in the transversal direction,
 $\partial \D/ \partial y=\partial \D/\partial z=0$.
Moreover, as the diffusion modes cannot propagate
through the clean metallic leads,  the diffusion propagator
must vanish at the contacts  to the latter.

The action (\ref{e55}) can be viewed as a differential operator
acting in the $2\times 2$ space spanned by  the fields $w$ and
$\bar{w}$. We need to invert this operator to obtain their
correlation  function.
\begin{mathletters}
\label{prop1,prop2,prop3}
\begin{eqnarray}&&
\langle w(x,\epsilon_1,\epsilon_2)
\bar{w}(x',\epsilon_3,\epsilon_4)
\rangle= \nonumber \\&&
2(2\pi)^2\delta(\epsilon_1-\epsilon_4)
\delta(\epsilon_2-\epsilon_3) \D(x,x',\epsilon_1-\epsilon_2)
\label{prop1} \, ,\\&& \langle
w(x,\epsilon_1,\epsilon_2)w(x',\epsilon_3,\epsilon_4)\rangle=-(2\pi)^3\delta(\epsilon_1-\epsilon_4)
\delta(\epsilon_2-\epsilon_3) \cdot
\nonumber  \\&& g\int dx_1
\D_{\epsilon_1-\epsilon_2,x,x_1}\nabla F_{\epsilon_2,x_1} \nabla
F_{\epsilon_1,x_1} \D_{\epsilon_2-\epsilon_1,x_1,x'} \label{prop2}
\,\, ,\\&& \langle \bar{w}(x,\epsilon_1,\epsilon_2)
\bar{w}(x',\epsilon_3,\epsilon_4) \rangle=0 \,\, \label{prop3},
\end{eqnarray}
\end{mathletters}
were the dimensionless conductance $g=\hbar \nu D$.
Note that as a result of the  Keldysh rotation, eq.(\ref{rot1}),
the correlation function of $\langle\bar{w}\bar{w}\rangle$
(eq.(\ref{prop3})) vanishes. Next we employ the action (\ref{e55})
to calculate various observables. For this purpose we will use the
generating functional, eq. (\ref{aa3}). To write it in terms of
the $Q$ matrices we note that the magnetic field enters the action
(\ref{e6}) through the ''long derivative'' $\partial=\nabla+i[{\bf
a}^\alpha\gamma_\alpha,\,\,\, ]$, where $e {\bf a}/c$ is the
vector potential. This  follows from the gauge transformation
\cite{Lerner} $\Psi \rightarrow U_A\Psi $ which allows to
eliminate the vector potential from the action. Provided that the
spectrum of the electrons can be approximated as linear,  one
chooses $ \nabla U_A= -iA \tau U_A$. As a result the generating
functional (\ref{aa3}) can be written as
\begin{eqnarray}&&
\label{gen2}
Z[{\bf a}]=\int {\cal{D}}Q \exp\left(-\frac{\pi\nu\hbar}{4}\Tr[D \left( \partial Q \right)^2 +4i \hat{\epsilon} Q]\right)
\end{eqnarray}
As a simple example for an observable we consider the mean
electron current through the bridge (at a vanishing magnetic
field), which is given by

\begin{eqnarray}&&
\label{f1}
I=\frac{-e }{2i}\frac{\delta Z[{\bf a}]}{\delta {\bf a}_2}\bigg|_{{\bf a}_2=0;{\bf a}_1=0} \,\, .
\end{eqnarray}
The corresponding leading  diagram for the average current is
shown in Fig.(\ref{fig9})

\begin{minipage}{3.1in}
\begin{figure}
\epsfysize=.2 in
{\makebox(130,-30){${G^R}$}}
{\makebox(270,-50){${G^K}$}}
\centerline{\epsfbox[0 65 250 100]{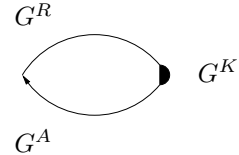}}
{\makebox(130,70){${G^A}$}}
\caption{The leading diagram for the  mean value of the electron current in the diffusion approximation.
The shaded semi-circle represents the Keldysh component of the Green function.}
\label{fig9}
\end{figure}
\end{minipage}

\vspace {1.5cm}
Employing eq. (\ref{gen2}) we find an expression for the d.c. current
\begin{eqnarray}&&
\label{ff1}
I=e\pi \nu\hbar D \Tr\{\big(Q\nabla Q-\left(\nabla Q\right) Q\big)\gamma_2\} \,\, .
\end{eqnarray}
Taking the saddle point solution for $Q$, eq. (\ref{e23}), one obtains
\begin{eqnarray}&&
I_{{\rm Ohm}}=G_{{\rm Ohm}}V \,\, ,
\end{eqnarray}
which is Ohm's law; for diffusive systems of length $L$
and cross-section $\cal{A}$ the low frequency conductance is equal to
\begin{eqnarray}&&
\label{ohm}
G_{{\rm Ohm}}=\sigma \frac{{\cal{A}}}{L}\,\, ,
\end{eqnarray}
where the conductivity $\sigma$ is related to the diffusion
coefficient and the thermodynamic density of states through the
Einstein relation
\begin{eqnarray}&&
\sigma =e^2 \nu D\,\, .
\end{eqnarray}
\section{Current Noise in Disordered Junctions}
\label{sec_3}
Besides the average value of the current one may study its
fluctuations  in time \cite{Kogan_book,Shoelkopf}. As far as the
second moment is considered, one is often concerned with the
symmetrized correlation function \cite{LL}
\begin{eqnarray}
\S(t,x,t',x')=\!\frac{1}{2}\!\langle \langle \hat{I}(t,x)\hat{I}(t',x')+\hat{I}(t',x')\hat{I}(t,x)
\rangle\rangle.
\end{eqnarray}
In a steady state  $S$ is a function of the difference between its
respective arguments. In equilibrium this quantity is  related to
the linear conductance  by the Fluctuation-Dissipation Theorem
(FDT)
\begin{eqnarray}&&
\label{e34}
\S^{eq}(\omega)= \hbar\omega \coth\left (\frac{\hbar\omega}{2T}\right)G(\omega)\, ,
\end{eqnarray}
where $G(\omega)$ is the frequency dependent conductance. Note
that the choice of the microscopic parameters ( such as impurity
strength and concentration ) enter only through the diffusion
coefficient, hence the conductance. The relation, eq.
$(\ref{e34})$,  between the conductance and current noise is
universal at equilibrium and does not depend on the microscopic
details of the system at hand. Out of equilibrium there is no
fundamental expression governing the  relation between conductance
and the spectral function of the noise, thence one needs to
calculate them separately.

To calculate a two-operator  correlation function, one  may choose
the time indices  to relate to  different branches of the Keldysh
contour, and consequently represent the correlator as a Keldysh
time-ordered expression, making it  possible to apply Wick's
theorem. Formally, $\langle\hat{A}(t)\hat{B}(t')\rangle=\langle
T_c \hat{A}(t^+)\hat{B}(t'^-)\rangle $, where $\hat{A}, \hat{B}$,
are arbitrary operators and $T_c$ denotes the time ordering
operator on the Keldysh contour. The symmetrized current-current
correlation function (at a vanishing magnetic flux) can be
represented as
\begin{eqnarray}
\S(x,t;x',t')=-\frac{e^2}{4}\frac{\delta^2 Z[{\bf a}]}{\delta {\bf a}_2(x,t)\delta {\bf a}_2(x',t')}\Bigg|_{{\bf a_1}=0;{\bf a}_2=0}
\,\, .
\end{eqnarray}
Here we have used  the identity $Z[{\bf a}_1; {\bf a}_2=0]=1$.
After functional differentiation  one obtains the following equation
\begin{eqnarray}&&
\label{e7}
\S(x,t;x',t')=\frac{e^2\pi\hbar\nu D}{4}
\bigg\langle I^D_{x,t;x',t'} -\frac{\pi\nu D}{2}M_{x,t} M_{x',t'}\bigg\rangle_0 \,\,,
\end{eqnarray}
where the subscript $0$ denotes  averaging with the action (\ref{e6});
we have also introduced
the notation
\begin{eqnarray}&&
\gamma_1=\left(\matrix {1 & 0\cr 0 & 1 \cr}\right) , \;
\gamma_2=\left(\matrix {0 & 1\cr 1 & 0 \cr}\right), \nonumber \\&&
I^D_{x,t,t'}=\Tr\!\bigg\{\!Q_{x,t,t'}\gamma_2Q_{x',t',t}\gamma_2-\delta_{t,t'}Q_{x,t,t'}Q_{x',t',t}\bigg\}\delta_{x,x'}  \nonumber \\&&
M(x,t)=\Tr\bigg\{\int dt_1\bigg[Q(x,t,t_1)\nabla Q(x,t_1,t)-
\nonumber \\&&
(\nabla Q(x,t,t_1)) Q(x,t_1,t) \bigg]\gamma_2\bigg\} \,\, .
\end{eqnarray}
One may note the correspondence between the expression (\ref{e7}) and
the direct diagrammatic analysis published earlier \cite{Altshuler-Levitov}
(cf. Fig. \ref{fig8}).

\vspace{1in}
\begin{minipage}{3.1in}
\begin{figure}
\epsfysize=.3 in
\centerline{\epsfbox[0 70 750 150]{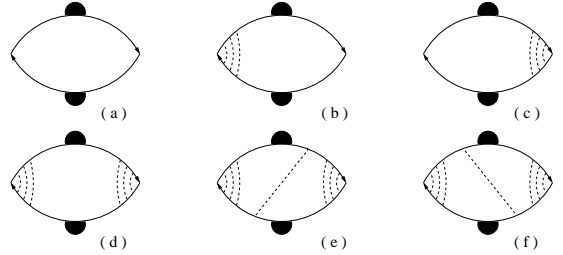}}
\vspace{0.5 in}
\caption{The leading diagrams for the current noise;
the shaded semi-circle represents  the single particle distribution function;
the dashed lines represent contraction of impurity lines
(i.e., ensemble averaging)}
\label{fig8}
\end{figure}
\vspace{0.1 in}
\end{minipage}

The first (diamagnetic) term in eq. (\ref{e7}) is
local in space (i.e. decaying on distance of the order of the
elastic mean free path) and corresponds to the bare diagram
(\ref{fig8}a), which describes the fluctuations of the
distribution function at a given point. The non-local part
corresponds to the diagrams (\ref{fig8}b-f), which describe the
diffusive propagation of current fluctuations throughout the
sample. It is convenient to define a local effective noise
temperature as
\begin{eqnarray}&&
\label{g7}
T_{noise}(\omega,x)=\frac{1}{4}\int d\epsilon
[1-F_{\epsilon+\omega,x}F_{\epsilon,x}] \,\, ,
\end{eqnarray}
such that at equilibrium the  noise can be cast as purely
``thermal'' :
\begin{eqnarray}&&
S^{eq}=2T_{noise} G \, .
\end{eqnarray}
One may work out (see  Appendix \ref{ap_1}) the explicit form of the
current-current correlation function:
\begin{eqnarray}&&
\label{e30}
S(x,x',\omega)=2 e^2g
\bigg[
T_{noise}(x,\omega) \delta(x-x')+ \nonumber \\&&
\pi g\big[\nabla\nabla'\D_{x,x',\omega}\!T_{noise}(x,\omega)+
\nabla\nabla'\D_{x,x',-\omega}\!T_{noise}(x',\omega)\!+ \nonumber \\&&
\pi g\int\!dx_1\nabla\nabla_1\D_{x,x_1,\omega}\nabla'\nabla_1
\D_{x_1,x',-\omega}\!T_{noise}(x_1,\omega)\big]
\bigg]\,.
\end{eqnarray}
The same expression was independently  derived by Nagaev recently
\cite{Nagaev3}.

Although eq.(\ref{e30}) is generally quite complicated, one may
study certain  limiting cases thereof.

In the low frequency limit  ( $\omega \ll E_{Th}$)  the
correlation function does not depend on the spatial coordinates
due to particle conservation. To study the temperature and
frequency dependence it is convenient to integrate  eq.
(\ref{e30}) over the volume of the system , which leads to
\cite{Altshuler-Levitov}
\begin{eqnarray}&&
\label{e25}
\S(\omega)=\frac{1}{6}[4\S^{eq}(\omega)+\S^{eq}(\omega+eV)+\S^{eq}(\omega-eV)]\,\, .
\end{eqnarray}

In the high frequency limit ($\omega \gg E_{Th}$) the behavior is
qualitatively different. The  "cross-coordinate'' noise spectrum
(cf.  eq.(\ref{e30})) in this situation is a decreasing function
of the distance between cross-section ($x,x'$), through which the
current fluctuation are studied. If the distance $l_{x,x'} \equiv
|x-x'|$ is much larger than the
 typical diffusion length $\sqrt(D/\omega)$,
the correlation function (\ref{e30}) practically vanishes.

We  stress that while the electrons here are assumed to be
non-interacting, {\it the low frequency limit} of equation
(\ref{e25}) {\it is  valid for interacting electrons as well } (in
the limit of  low electron-electron collision rate) , and is in
good agreement with experiment \cite{Shoelkopf}. To see this one
may follow arguments similar to those given by
Nagaev\cite{Nagaev3,Nagaev2}. Let us assume that it is possible to
describe the kinetics of {\it interacting} electrons by a
Kogan-Shulman type set of equations
\begin{eqnarray} &&
\label{z5}
\delta j=-e D\nabla \delta n+\sigma(\omega)\delta E(\omega)+\delta j^{ext} \nonumber \, \, ,\\&&
i\omega e\delta n+ \nabla \delta j =0 \,\, ,\nonumber\\&&
\nabla \delta E=4\pi e \delta  n \,\, .
\end{eqnarray}
Here the fluctuations of the electric current density, $\delta j$,
are related to the fluctuations of the particle density , $\delta
n$, and the electric field, $\delta E$, in a self-consistent field
approximation.  For simplicity we have assumed here that the
elastic relaxation time and the electron density of states are
constant in  the vicinity of the Fermi surface. Furthermore, we
have ignored the contribution to the resistance associated with
elastic scattering at the contacts to the leads. If this is not
the case, the eq.(\ref{z5}) should be modified correspondingly
\cite{Gomila}.

 The ``random source'' $\delta
j^{ext}$ is generated by  short-range scattering events of
conducting electrons on the impurities.  Motivated by the fact that
scattering events
on  scales  larger than the mean free path are uncorrelated,  we model
the random source correlation
function by
\begin{eqnarray} &&
\label{bb11} \langle \delta j^{ext}_\alpha(r) \delta j
^{ext}_\beta(r')\rangle_\omega= 4 \pi
\sigma(\omega)\delta(r-r')T(r,\omega)\, .
\end{eqnarray}
In the  low frequency limit the continuity equation guarantees
that the current-current correlation function is independent of
the choice of  cross-sections, i.e., it is  constant as function
of its coordinates. To find this constant it is convenient to
integrate the current correlation function with respect to its
coordinates over the volume of the  sample. In our model there are
no fluctuations of the electron density and  electrostatic
potential in the leads. This implies that   the gradient term
disappears after the volume integration, i.e.,  we rederive the
non-interacting  result, eq.(\ref{e25}). Given the fact that
neither the diffusion propagator nor the effective noise
temperature entering eq.(\ref{e30})  depend on the phase of the
electron wave-function, we conclude that the  current noise in
diffusive systems   is unaffected by quantum decoherence processes
in the system\cite{Nagaev1,Nagaev2,deJong_Beenakker} to  leading
order in $1/g$.

\section{current fluctuations of interacting electrons}
\label{sec_4}
So far we have considered the physics of  non-interacting electrons.
In the present section we study how the electron-electron interaction affects
the current through a diffusive junction as well as the current  fluctuations,
all this out of equilibrium.
One may note that the conductance ( {\it including }   weak localization
corrections) is insensitive to the shape of the electron distribution function.
By contrast, the double-step shape of the distribution function
is crucial as far as shot-noise is concerned.
This provides us with a motivation to consider the effect of electron-electron interaction on the single-electron distribution function, hence the noise.
Since the electrons are interacting particles, one should include the (electron
density dependent)
electric field in the kinetic operator,
supplemented by a self-consistency condition, eq.(\ref{z5}).
This leads to the appearance of a smooth electric field in the
constriction which, on its own
(as we have already seen in the Section \ref{sec_1}),
does not give rise to  any effect on the low frequency noise,
but may modify the higher frequency spectrum of the current noise.
This case has been studied  extensively applying the Shulman-Kogan approach
\cite{Shulman} to various  physical system \cite{Nagaev1,Nagaev2,Nagaev3}.
Briefly, the presence of interaction defines a
characteristic $RC$ time scale.
Depending on the ratio between the measured frequency and the inverse of
this time scale, fluctuations of the distribution function propagate
(or do not propagate) through the entire system
( over time scales $ \sim 1/\omega$ ).
In the low frequency limit, fluctuations of the distribution function
{\it do}  propagate throughout the entire system  (over time scales
$ \sim 1/\omega$),
in which case  current-current correlators are independent
of the spatial coordinates and do not discriminate between
interacting and non-interacting electrons.
The value of the shot noise is then given by
\begin{eqnarray}
\label{shot1}
\S(0)=\frac{eI}{3} \,\, .
\end{eqnarray}
By contrast, at high frequencies ( larger then the inverse $RC$ time)
interactions affect the frequency and the space dependence
of the current noise \cite{Nagaev2,Likharev}.

However, in addition to the creation of a smooth electric field,
electron-electron interactions may lead to inelastic scattering.
The relative importance of such a scattering process
is determined by the ratio between the sample size $L$ and the  inelastic mean free path $l_{in}$.
When the inelastic length is much larger than the system size,
 electron-electron scattering leads to a small positive correction
to the noise given by eq.(\ref{e25}) (cf. Ref.(\cite{Nagaev95})).
For systems with a short relaxation length
($l_{in} \ll L$), the distribution function can be approximated
by a quasi-equilibrium one.
The latter implies that each piece of the conductor possesses
its own effective temperature and electro-chemical potential.
 The contributions of  the individual pieces are then
added independently.
The local quasi-equilibrium conditions lead to the ``smearing''
of the double-step distribution function obtained for non-interacting electrons
out of equilibrium, eq.(\ref{dist1}),
and consequently to the enhancement of the current noise
(or in other words to a partial suppression of the electron-electron  correlation) \cite{Nagaev95,Kozub}
\begin{eqnarray}
\S(0)=\frac{\sqrt{3}}{4}eI \,\, .
\end{eqnarray}

We now focus on the  limit of large inelastic length, and study
the effect of electron-electron interaction {\it beyond} the
applicability of the kinetic equation;  we evaluate the
interaction corrections to the current noise. While this
contribution is small in magnitude,  for system with effective
dimensionality $d \le 2$   it has strong (singular) dependence on
temperature and voltage, contrary to the large ( but practically
constant ) non-interacting part.

We take  the function $F$ to be that of the non-interacting gas
(in general out of equilibrium), eq.(\ref{dist1}). Corrections to
the equilibrium conductance ( and noise) due to the interplay
between interaction and disorder have originally been discussed by
Altshuler and Aronov \cite{AA,AAL} and in two-dimensional settings
are given by
\begin{eqnarray}&&
\label{a60}
\delta G=\frac{G}{2\pi^2 g}\ln\left(\frac{T\tau}{\hbar}\right) \,\, .
\end{eqnarray}

One may note that unlike the non-interacting case (cf. eq. (\ref{ohm})),
the conductance due to the Altshuler-Aronov term depends strongly
on temperature,  i.e. on the smearing of the distribution function.
In the non-equilibrium problem the shape of the distribution function
is markedly different from the Fermi-Dirac distribution function, which
motivates us to study  analogous, out-of-equilibrium corrections to the current {\it and} to the noise.
In order to do so we incorporate  the Coulomb interaction into the sigma model.
The action of the interacting electrons may be written as
\begin{eqnarray}&&
\label{r1}
S_{total}[\Psi]=S_0[\Psi]+S_{int}[\Psi] \,\, ,
\end{eqnarray}
where $S_0[\Psi]$ is given by eq. (\ref{aa10}) and
the interaction part of the action can be described as
\begin{eqnarray}&&
\label{bb1}
S_{int}[\!\Psi]\!=\!-\frac{1}{2}\!\sum_{i=1}^2\!\int\!d{\bf\!r}d{\bf\!r'}dt\rho_i(r,t)\sigma_i^{(3)}V_0(\!r-\!r'\!)\rho_i(r'\!,t).
\end{eqnarray}
Here $\rho$ is the electronic density, $\rho_i=\Psi^\dagger_i\Psi_i$,
while $V_0$ represents the interaction potential.
One may introduce a new auxiliary bosonic field
\begin{eqnarray}&&
\tilde{\Phi}=\left(
\begin{array}{l}
\tilde{\phi}_1 \\
\tilde{\phi}_2
\end{array} \right)\,\, ,
\end{eqnarray}
which  decouples the interaction in the particle-hole channel, yielding
\begin{mathletters}
\begin{eqnarray}&&
\label{act1,act2,act3}
iS_{total}=iS[\tilde{\Phi}]+iS[\tilde{\Phi},\tilde{Q}] \, ,
\label{act1} \\&& iS[\tilde{\Phi}]=i\hbar\Tr\{\tilde{\Phi}^T
V^{-1}_0\sigma^{(3)}\tilde{\Phi}\}  \, , \label{act2} \\&&
iS[\tilde{\Phi},\tilde{Q}]=-\frac{\pi\hbar\nu}{4\tau}\Tr\{\tilde{
Q}^2\}\!+\nonumber \\&&
\Tr\ln \Big[\hat
G_0^{-1}\sigma^{(3)}\!+\frac{i\tilde{Q}\sigma^{(3)}}{2\tau}\!+\tilde{\phi}_{\alpha}\tilde{\gamma}^{\alpha}
\Big].  \label{act3}
\end{eqnarray}
\end{mathletters}
A major difficulty now arises.
The study of an electron gas
{\it at equilibrium }relies on the fact that the correlation function of the fluctuations of the electric potential in the sample can be expressed through the polarization operator of the gas. This is not the case out of equilibrium.
To find such a correlation function in the non-equilibrium case is a demanding  task, which, in its general formulation, is yet to be solved.
Formally the difficulty arises because the conventional RPA approximation has to be modified.
Physically this is related to the fact that screening of the interaction for the non-equilibrium gas differs from the standard equilibrium problem.
We circumvent  this difficulty by considering a toy model of an instantaneous
 short-range interaction:
\begin{eqnarray}&&
\hat{V_0}({\bf r}-{\bf r'};t-t')=\hbar\Gamma \delta({\bf r}-{\bf r'})\delta(t-t') \,\, .
\end{eqnarray}
The effective interaction strength $\Gamma$ is assumed to be
small:
\begin{eqnarray}&&
\label{small_interaction}
\Gamma\nu \ll 1\,\,.
\end{eqnarray}
The condition, eq.(\ref{small_interaction}), allows to calculate
corrections due to electron-electron  interactions perturbatively.
 From calculations done at equilibrium it is known that while this
model predicts the correct qualitative behavior for the electronic
conductance, it misses the double logarithmic correction to the
tunneling density of states. The different singularities in the
conductance and the tunneling-density-of-states have  been
attributed to the presence (absence) of an approximate gauge
invariance symmetry \cite{Fin2} in the former (latter). Gauge
invariant quantities are less sensitive to the details of the
electron-electron interaction. Since the current correlation
function is evidently  gauge invariant, we expect our model
interaction to yield the correct (singular) temperature and
voltage dependence, up to a non-universal numerical prefactor.

Following the rotation, eq.(\ref{rot1}), of the interaction part
of the action, eq.(\ref{act2}), in  Keldysh space, it reads
\begin{eqnarray}&&
\label{e51}
iS[\Phi]=i\Gamma \Tr\{\Phi^T({\bf r},t) \sigma^1 \Phi({\bf r},t)\}\,\, .
\end{eqnarray}

Next one performs a gradient expansion of eq.(\ref{act3}), treating
the  field $\phi$ under the logarithm as a small perturbation, which
yields
\begin{eqnarray}&&
\label{de52}
iS[\Phi,Q]=
-\frac{\pi\nu}{4}\Tr\{D (\nabla Q)^2\!\!-\!4i (\phi_\alpha\gamma^\alpha\!+\!\hat{\epsilon})Q \}\,\,.
\end{eqnarray}
The correlation function of the fluctuations of the electric
potential, corresponding to eq.(\ref{e51}), is
\begin{eqnarray}&&
\label{e52} \langle
\phi_\alpha(r,\omega)\phi_\beta(r',-\omega)\rangle =-i\Gamma
\delta(r-r')\sigma^{(1)}_{\alpha,\beta} \, \, ,
\end{eqnarray}
and the correlation function of $w$ and $\bar{w}$ are given by
eq.(\ref{prop1,prop2,prop3}). Let us now use this model to
calculate the interaction  correction to the d.c. value of
electron current. Employing eq.(\ref{f1}) we obtain

\begin{eqnarray}&&
\label{f2}
\delta I^{ee}=\frac{e\pi \nu D (i\pi \nu)^2}{4}\times \nonumber \\&&
\langle \Tr\{(Q\nabla Q-\nabla Q Q)\gamma_2\}\Tr\{\phi_\alpha\gamma^\alpha Q\}
 \Tr\{\phi_\beta \gamma^\beta Q \}\rangle_0 \, \, ,
\end{eqnarray}

where the subscript $0$ means that we perform averaging employing
the correlators of eqs.(\ref{e52}) and (\ref{prop1,prop2,prop3}).

The analogous correction to the current noise is given by 
\begin{eqnarray}&&
\label{cor1}
\delta\S^{ee}(0)\!=\!\frac{e^2\pi\!g(i\pi\nu)^2}{4}
\bigg<\![I^D_{x,t,t'}\!\delta_{x,x'}\!-\!\frac{\pi\!\nu\!D}{2}\!M_{x,t}
M_{x',t'}]\!\!\times\nonumber \\&&\Tr\{\phi_\alpha\gamma^\alpha
Q\}\Tr\{\phi_\beta \gamma^\beta Q \}\bigg>_{0}.
\end{eqnarray}

So far our analysis was general and applied to diffusive
conductors of any dimensionality. From this point on we are going
to focus on quasi-two-dimensional systems and study eqs.(\ref{f2})
and (\ref{cor1}) for such geometries.  Evaluating the expression
of eq.(\ref{f2}) (see Appendix \ref{ap_2}) we obtain a correction
to the electron current. The strongest functional dependence of
that correction on temperature (or voltage) is obtained when the
parameter $\zeta$=\max$(eV,kT)$ is larger than the Thouless
energy, i.e. \cite{cutof} $\zeta \gg E_{Th}$ . Our analysis then
yields
\begin{eqnarray}&&
\label{f8}
\delta I^{ee}= \frac{I_{{\rm Ohm}}}{2\pi^2g} \ln(\zeta\tau)  \,\, .
\end{eqnarray}

Eq.(\ref{f8}) for the effect of interaction on the current  at
$d=2$ is valid to the leading logarithmic correction. It features
both the voltage and the temperature in a symmetric way, namely as
cutoffs which govern the singularity. For  high temperatures, $T
\gg eV$, the singularity is determined by $T$, which is consistent
with the celebrated  Altshuler-Aronov correction to the
conductance.

We now turn to the calculation of the interaction correction to
the current noise eq. (\ref{cor1}).

For  frequencies smaller than the Thouless frequency the
current-current correlation function does not depend on the choice
of cross-section. Moreover, we also assume that $T$ is much larger
than the Thouless energy\cite{cutof}. To calculate the  correction
to the current noise due to electron-electron interaction one
expands eq.(\ref{cor1}) in the fields $w,\bar{w}$. After averaging
over these fields (see appendix \ref{ap_1})  we finally obtain
\end{multicols}
\top{-3cm}
\begin{eqnarray}&&
\label{e29} \delta\S^{ee}(0)\!=\!\frac{G_{{\rm Ohm}}}{6\pi^2g}\!
\bigg[\!2T\ln\left(\!\frac{\zeta\tau}{\hbar}\!\right)
\!+\!eV\coth\left(\!\frac{eV}{2T}\right)\bigg(
\!\ln\left(\!\frac{\zeta\tau}{\hbar}\!\right)\!+\!\ln\!\left(\!\frac{T\tau}{\hbar}\!\right)\!\bigg)
\bigg].
\end{eqnarray}
\bottom{-3cm}
\begin{multicols}{2}
Eq. (\ref{e29}) is the main result of this section. Considering
the asymptotic high temperature behavior of eq.(\ref{e29}), one
notes  that  $\delta\S(0)$ is given to leading order by the
expression for  Nyquist noise. Corrections to the noise in this
limit are related to the interaction corrections to  the
conductance \cite{AA} (the latter  is related to the Nyquist
noise through the FDT), namely
\begin{eqnarray}&&
\label{corr3} \delta \S^{ee}(0)=\frac{G_{{\rm Ohm}}}{\pi^2g}T\ln\left(\frac{T\tau}{\hbar}\right),
\end{eqnarray}
Eq.(\ref{corr3}) is related  to the correction to the mean
current, eq.(\ref{f8})).

In the large voltage limit, the shot component of the noise is
dominant, but interaction correction to it are still determined by
the temperature
\begin{eqnarray}&&
\label{r2} \delta \S^{ee}=\frac{G_{{\rm Ohm}}}{6 \pi^2g}e|V|\ln\left(\frac{T\tau}{\hbar}\right).
\end{eqnarray}
Similarly to previous high temperature limit, here too the
interaction correction to the noise is negative (cf. \ref{corr3}).
As we can see the results for the current (\ref{f8}) and for the
current noise (\ref{r2}) are no more simply related  to each
other.

The result we have obtained can be qualitatively  understood in
the following way. The charge transfer through disordered junction
is a stochastic process. For non-interacting electrons the
transfer of  the charged particles is described by a time sequence
with  a binomial distribution function \cite{Levitov_Lee}.  The
mean current has to do with the mean charge transferred through a
cross-section over a given time interval, while the second
cummulants  of the current noise is related to the dispersion of
the distribution function. Electron-electron interactions render
the electron motion more correlated. This affects the dispersion
more strongly than it affects the average value of the
distribution, implying that the ongoing random process is no
longer binomial. The deviation from the binomial distribution,
though, is small in the inverse dimensionless conductance.

  Finally we  would like to  discuss in brief  possible experimental
implications of our results. The effect  predicted here can, in
principle, be tested  experimentally in  samples where the
conventional Altshuler-Aronov suppression of the conductance is
observed. The non-equilibrium character of the quantities
considered  raises the  question  whether other mechanisms (such
as inelastic electronic collisions) which might modify the noise
spectrum as well, would not mask our effect. We thus would like to
point out a qualitative difference between the interaction
corrections to the noise discussed here, and  corrections that
arise because of the  inelastic length being finite
\cite{Nagaev95,Kozub}. Normally the electron inelastic length
increases as the temperature and/or the  applied bias decrease.
For this reason  corrections  associated with the ratio $L/l_{in}$
being finite diminish as one lowers the bias (temperature).
Indeed, when the inelastic length is determined by large momentum
transfer
 the correction to the shot noise due to {\it
inelastic collisions} is given by \cite{Nagaev95,Kozub}
\begin{eqnarray}&&
\label{dS_inelastic} \delta S^{in}=
\frac{eI}{3}\frac{12\pi^2}{40320}\frac{\kappa_3}{p_F}\frac{(eVL)^2}{D
\epsilon_F}=0.02 \cdot e^2 I\frac{L^2}{l_{in}l}\,\, .
\end{eqnarray}
Here $\kappa_3=\sqrt(4\pi e^2\nu)$ is the inverse screening length
in three dimensions. The second equality in
eq.(\ref{dS_inelastic}) makes use of the expression for the
three-dimensional inelastic length (large momentum transfer),
given by
\begin{eqnarray}&&
l_{in}=\frac{64}{\pi^2}\frac{\epsilon_F^2}{(eV)^2}\frac{1}{\kappa_3}\,\,.
\end{eqnarray}
This expression is to be used for  quasi-two-dimensional
conductors (metallic films). While the leading term, eq.
(\ref{shot1}), is linear in the voltage, the corrections are
proportional to its second power, and therefore can be neglected
for small enough voltages. The corrections we have found,
eqs.(\ref{corr3},\ref{r2}), behave in the opposite way. They
become increasingly  more pronounced as one lowers the voltage and
the temperature. For a quantitative estimate we consider a
metallic film of  length $L=10^{-4}\rm{cm}$ and of thickness
$10^{-6}{\rm cm}$, having  sheet resistance of $10 \,
\Omega/$square.  The diffusion coefficient is taken to be $D=10^3
\rm{cm}^2 \rm{sec}^{-1}$, the elastic mean free time --
$\tau=10^{-13}\rm{sec}$ and the Fermi energy $\epsilon_F=0.1eV$.
We consider a voltage bias of $V=10\mu V$. The current noise in
this range of parameters is of the order of $10^{-24}{\rm
Coulomb}\, {\rm Ampere}$. The corrections to the current noise due
to  inelastic collisions are in this case of  order
$10^{-29}\rm{Coulomb}\, \rm{ Ampere}$, while the corrections due
to the quantum interactions are of  order $10^{-26}\rm{Coulomb}\,
\rm{Ampere}$.

{\bf Acknowledgments}.
We acknowledge helpful discussions with A.~Kamenev, D.~E.~Khmel'nitskii,
L.S.~Levitov, M.~Reznikov, D.~Prober and R.J.~Schoelkopf.
This work was supported  by the
U.S.-Israel Science Foundation (BSF), the German-Israel Foundation (GIF),
the ISF founded by the Israel Academy of Sciences and Humanities-Centers of
Excellence Program, and by the DIP foundation.

\appendix
\section{Noise of non-interacting electrons}
\label{ap_1}
In this Appendix we derive eq.(\ref{e30}). One may note that there
are two types of terms in eq.(\ref{e7}). The "diamagnetic" part
(arising from the ${\bf a^2}$ term in the action)  is proportional
to the dimensionless conductance, and therefore it is sufficient
to consider its value at the saddle point:
\begin{eqnarray}&&
\I_0^D(x,t)=\frac{4}{(2\pi)^2}\int d\epsilon d\Omega \exp(i
\Omega t) \nonumber \\&&
[1-F(x,\epsilon-\Omega)F(x,\epsilon)].
\end{eqnarray}
The $\langle \M\M \rangle$ correlator  is proportional to the
square of the  dimensionless conductance. It   contains  both
reducible and irreducible contributions to the current-current
correlation. In order to  extract the   contribution of this
correlator to the second order {\it cummulant}  of the current
(irreducible contribution), we will need to expand it up to second
order in $w, \bar{w}$. Expanding $M$ to first order in these
variables\cite{wlc} one obtains
\end{multicols}
\top{-2.8cm}
\begin{eqnarray}&&
\label{f30}
\M_1(x,t)=
\frac{2}{4\pi^2}\int \exp(i\Omega t)d\epsilon d\Omega \nonumber \\&&
\bigg[\nabla\bar{\omega}(x,\epsilon,\epsilon\-\Omega)\!+
\nabla\omega(x,\epsilon,\epsilon-\Omega)-
\nabla\bar{\omega}(x,\epsilon,\epsilon-\Omega)F(x,\epsilon)F(x,\epsilon-\Omega)+ \nonumber \\&&
\nabla F(x,\epsilon-\Omega)F(x,\epsilon)\bar{\omega}(x,\epsilon,\epsilon-\Omega)+\nabla F(x,\epsilon)F(x,\epsilon\!-\!\Omega)\bar{\omega}(x,\epsilon,\epsilon\-\Omega)\bigg]
\end{eqnarray}
( Here and later on the subscript $k$ in the $\I_k$ and $\M_k$
denotes the power of fluctuating field up to which it is
expanded). Now one needs to average the product of $M_1$'s over
the  fluctuating fields
\begin{eqnarray}&&
\label{f31} \langle \Tr\{\left(Q\nabla Q-\nabla Q Q\right)\gamma_2
\}_{x,t} \Tr\{\left(Q\nabla Q-\nabla Q Q\right)\gamma_2
\}_{x',t'}\rangle _{w,\bar{w}}=\nonumber \\&& \frac{2}{\pi^2} \int
d\epsilon d\Omega e^{i\Omega(t-t')} \bigg[ \nabla
\nabla'\D_{\Omega}(1-F_{\epsilon,x'}F_{\epsilon-\Omega,x'})+
\nabla \nabla'\D_{-\Omega}(1-F_{\epsilon,x}F_{\epsilon+\Omega,x})+
\nonumber \\&& \pi \nu D \int d x_1 \nabla \nabla_1\D_{\Omega}
\nabla'\nabla_1\D_{-\Omega}
(1-F_{\epsilon,x_1}F_{\epsilon-\Omega,x_1}) \bigg]\,\, .
\end{eqnarray}
Combining  eq.(\ref{f30}) and eq. (\ref{f31}) with eq.(\ref{e7})
one obtains the expression for the current correlation function
through different cross sections at a finite  frequency (for
which the diffusion approximation holds)
\begin{eqnarray}&&
\label{ap1}
\S(x,x',\omega)=e^2\pi\nu D \int \frac{d\epsilon}{2\pi}
\bigg[
[1 - F(x,\epsilon) F(x,\epsilon-\omega)] \delta(x-x')+
\nonumber \\&&
\pi \nu D \bigg(
 \nabla \nabla' \D_{x,x'\!,\omega}[1 - F_{x,\epsilon} F_{x,\epsilon - \omega}]
+
\nabla \nabla' \D_{x,x',-\omega}[1 - F_{x',\epsilon} F_{x',\epsilon - \omega}] +
\nonumber \\&&
\pi \nu D
\int dx_1
\nabla \nabla_1 \D_{x,x_1,\omega} [1 - F_{x_1,\epsilon} F_{x_1,\epsilon-\omega}]
 \nabla_1 \nabla' \D_{x_1,x',-\omega}\bigg)
\bigg]\,\, .
\end{eqnarray}
\begin{multicols}{2}
The last result can be re-expressed in terms of an effective
electrons temperature (eq. (\ref{g7})), leading to eq.(\ref{e30}).

\section{Corrections to the d.c. current}
\label{ap_2}
In this appendix we derive  eq.(\ref{f8}), i.e. the interaction
corrections to the disorder-averaged value of the d.c. current.
One first expands the operator $\M$ up to  second order in the
fluctuating fields
\end{multicols}
\top{-3.2cm}
\begin{eqnarray}&&
\label{f3}
\M_2(x_1)= \int d[\epsilon]
\bigg[\nabla F_{x_1,\epsilon_1} \bar{w}_{x_1,\epsilon_1,\epsilon_3}w_{x_1,\epsilon_3,\epsilon_2} + w_{x_1,\epsilon_1,\epsilon_3}\bar{w}_{x_1,\epsilon_3,\epsilon_2}\nabla F_{x_1,\epsilon_2}\bigg] \,\, ,
\end{eqnarray}
where  $d[\epsilon]$ implies  integration over all $\epsilon$'s,
$\int d\epsilon/2\pi$. Next we evaluate the expansion up to second
order in the fields $w, \bar{w}$ of the quantity
\begin{eqnarray}&&
\label{f10}
\langle \Tr\{\phi_1\gamma_1 Q\}\Tr\{\phi_2 \gamma_2 Q \}\rangle_{\phi}=
i\Gamma\int d[\epsilon]
\bigg[\bar{w}_{x_2,\epsilon_{4},\epsilon_{5}}w_{x_2,\epsilon_{6},\epsilon_{6}+\epsilon_{4}-\epsilon_{5}}(\!F_{x_2,\epsilon_{5}}\!-\!F_{x_2,\epsilon_{4}}\!)\bigg] \,\, .
\end{eqnarray}
Using the expressions (\ref{f2},\ref{f3},\ref{f10}) we find that
the correction to the d.c. current is given by
\begin{eqnarray}&&
\label{f5} \delta I^{ee\!}=-\frac{i}{8}e\pi^3\nu^3\!D\Gamma
{\cal{A}}\!\int [d\epsilon] \nonumber \\&&
\bigg\langle
\bar{w}_{x_2,\epsilon_{4},\epsilon_{5}}
w_{x_2,\epsilon_{6},\epsilon_{6}+\epsilon_{4}-\epsilon_{5}}(F_{x_2,\epsilon_{5}}\!-\!F_{x_2,\epsilon_{4}})
(\nabla\!F_{x_1,\epsilon_1}\bar{w}_{x_1,\epsilon_1,\epsilon_3}w_{x_1,\epsilon_3,\epsilon_2}\!+\!
w_{x_1,\epsilon_1,\epsilon_3}\bar{w}_{x_1,\epsilon_3,\epsilon_2}\nabla
F_{x_1,\epsilon_2}) \bigg\rangle_{w,\bar{w}}.
\end{eqnarray}
After  averaging over the fields $w$ and $\bar{w}$ we find that,
to lowest order in the interaction amplitude and inverse
conductance, the correction is given by
\begin{eqnarray}&&
\delta I^{ee}=-ie\pi^3\nu^3 D \Gamma {\cal{A}}
\int d [\epsilon]dx_2 (F_{x_2,\epsilon_3}-F_{x_2,\epsilon_1})\nabla F_{x_1,\epsilon_1}
[\D_{x_1,x_2,\epsilon_3-\epsilon_1}\D_{x_2,x_1,\epsilon_3-\epsilon_1}-
\D_{x_1,x_2,\epsilon_1-\epsilon_3}\D_{x_2,x_1,\epsilon_1-\epsilon_3} ] \,\, .
\end{eqnarray}
Now we will use the fact that the diffusion propagator at
frequency $\Omega$ is a function of the difference of the spatial
coordinates; it decays on the scale $\sqrt{D/\Omega}$ which is
much smaller than the sample size, provided that $\hbar\Omega\gg
E_{\rm Th}$. This allows us to integrate over the relative
coordinate (assuming that the distribution function does not
change much on that distance),
\begin{eqnarray}&&
\label{f7} \delta I^{ee}=\!-e^2V D\Gamma{\cal{A}}\frac{(i \pi \nu
)^3}{2\pi L} \int
d[\delta\epsilon]d^2[q]\D^2[q,-\delta\epsilon]\coth\left(\!\frac{\delta\epsilon+eV}{2kT}\!\right)
\,\, .
\end{eqnarray}
\bottom{-3cm}
\begin{multicols}{2}
Integrating over $q$  in (\ref{f7}) we obtain
\begin{eqnarray}&&
\delta I^{ee} = -\frac{e\pi\nu\Gamma{\cal{A}}eV}{(2\pi)^3L}\int
\frac{d\delta\epsilon}{\delta\epsilon}\coth\left(\frac{\delta\epsilon+eV}{2kT}\right)
\,\, .
\end{eqnarray}
After integration over the energy $\delta\epsilon$ we finally find
the correction to the current which exhibits a logarithmic
singularity. The latter is smeared on a scale which  is determined
by the  maximal between the temperature and the voltage:
\begin{eqnarray}&&
\label{z26}
\delta I^{ee}=\frac{e^2\pi \nu \Gamma {\cal{A}} V}{L(2\pi)^3} \ln(\zeta \tau) \,\, .
\end{eqnarray}
Restoring the "universal" ( i.e., $\Gamma$-independent)
coefficient from the known equilibrium result we derive
eq.(\ref{f8}).

\section{Corrections to the Current Noise}
\label{ap_3}
In this Appendix we derive eq.(\ref{e29}). To find the interaction
correction to the current noise we use the fact that for
frequencies lower than the Thouless energy  the result is
independent of the cross-sections involved; it is  therefore
legitimate to integrate the current correlation function over the
volume of the system.

Employing  eq.(\ref{cor1}) we will need to expand the $\I^D$ and
the $\M\M$ terms in  $w$ and $\bar{w}$ up to a second and third
order respectively. We find that the  contribution coming from the
corrections to $\M\M$ vanishes after integration over space
coordinates (it contains terms of the type $\D_{x,L}-\D_{x,0}$).
To find the correction that arises from the ``diamagnetic term''
we need to expand the latter   up to  second order in  $w$ and
$\bar{w}$,
\end{multicols}
\begin{eqnarray}&&
\label{f21} \I^D_2(t,t')=\int
d[\epsilon]\exp(i((\epsilon_1-\epsilon_4)t+(\epsilon_3-\epsilon_2)t'))
\nonumber \\&& \bigg[F_{\epsilon_3}\!
F_{\epsilon_4}w_{\epsilon_1,\epsilon_2}
\bar{w}_{\epsilon_3,\epsilon_4}\!-\!\delta_{\epsilon_3,\epsilon_4}w_{\epsilon_1,\epsilon_5}
\bar{w}_{\epsilon_5,\epsilon_2}\!+\!
2\delta_{\epsilon_3,\epsilon_4}\!F_{\epsilon_2}\! F_{\epsilon_3}
w_{\epsilon_1,\epsilon_5}\bar{w}_{\epsilon_5,\epsilon_2}\!+\!
F_{\epsilon_1}\!
F_{\epsilon_2}w_{\epsilon_3,\epsilon_4}\bar{w}_{\epsilon_1,\epsilon_2}\!-\!
\delta_{\epsilon_1,\epsilon_2}\!w_{\epsilon_3,\epsilon_6}\bar{w}_{\epsilon_6,\epsilon_4}\!+\!
\nonumber \\&& 2\delta_{\epsilon_1,\epsilon_2}F_{\epsilon_1}
F_{\epsilon_4}
w_{\epsilon_3,\epsilon_6}\bar{w}_{\epsilon_6,\epsilon_4}\!-\!
\delta_{\epsilon_3,\epsilon_4}\bar{w}_{\epsilon_1,\epsilon_5}w_{\epsilon_5,\epsilon_2}\!+\!
2\delta_{\epsilon_3,\epsilon_4}F_{\epsilon_1}F_{\epsilon_3}
\bar{w}_{\epsilon_1,\epsilon_5}w_{\epsilon_5,\epsilon_2}\!
\nonumber \\&&
-\!\delta_{\epsilon_1,\epsilon_2}\bar{w}_{\epsilon_3,\epsilon_6}w_{\epsilon_6,\epsilon_4}\!+
2\delta_{\epsilon_1,\epsilon_2}F_{\epsilon_1} F_{\epsilon_3}
\bar{w}_{\epsilon_3,\epsilon_6}w_{\epsilon_6,\epsilon_4}\bigg]
\,\, .
\end{eqnarray}
Combining eq.(\ref{f21})  with eq.(\ref{f10}) and using the fact
that the  diffusion propagator decays on scales much shorter than
the system's size, we can  convert the  spatial integration over
the relative coordinate to an integral in Fourier space (
analogously to how it was done in Appendix \ref{ap_2}). We obtain
\begin{eqnarray}&&
\delta \I^D_{ee}(\omega) =4\Gamma{\cal{A}}\int d[\epsilon]d[q]d x
\Im\bigg\{ \D^2[q,\epsilon_1-\epsilon_2]\bigg\}(F_{x,\epsilon_1}-F_{x,\epsilon_2})
(2F_{x,\epsilon_1}F_{x,\epsilon_1-w}-1) \,\, .
\end{eqnarray}
Thus the correction (eq. (\ref{cor1}) to the noise is given by
\begin{eqnarray}&&
\label{ac1} \delta \S^{ee}(\omega) =\frac{e^2 D{\cal{A}}\Gamma
(\pi \nu)^3}{L^2} \int d^2[q] d[\delta\epsilon]
\Im\bigg\{\D^2[q,\delta\epsilon]\bigg\}
Y(\delta\epsilon,\omega,eV,T)\,\, ,
\end{eqnarray}
where we have  defined
\begin{eqnarray}&&
\label{cutof2} Y(\delta\epsilon,\omega,eV,T)=\int d[u]dx
(F_{u,x}-F_{u-\delta\epsilon,x}) (2F_{u,x}F_{u-w,x}-1) \,\, .
\end{eqnarray}
Integrating over  energy and ("center-of-mass") coordinate we find
that at zero frequency, $\omega=0$,
\begin{eqnarray}&&
\label{f22} Y(\delta\epsilon,0,eV,T)= \frac{L}{3}\bigg[
2kT\left(\coth\left(\frac{eV-\delta\epsilon}{2kT}\right)-\coth\left(\frac{eV+\delta\epsilon}{2kT}\right)\right)+\nonumber
\\&& eV\coth\left(\frac{eV}{2kT}\right) \left(
\coth\left(\frac{eV-\delta\epsilon}{2kT}\right)-\coth\left(\frac{eV+\delta\epsilon}{2kT}\right)
\right) -2eV\coth\left(\frac{eV}{2kT}\right)\coth\left(\frac{\delta\epsilon}{2kT}\right)
\bigg] \,\, .
\end{eqnarray}
Substituting eq.(\ref{f22}) into eq.(\ref{ac1}) we conclude our
calculation of the  interaction correction to the noise
\begin{eqnarray}&&
\label{f23} \delta \S^{ee}(0)=\frac{G\Gamma \pi \nu }{3 g (2
\pi)^3} \bigg[2 kT \ln( \zeta \tau)+eV
\coth\left(\frac{eV}{2kT}\right) \left( \ln(\zeta\tau)+\ln(kT\tau)
\right) \bigg] \,\, .
\end{eqnarray}
Restoring a $\Gamma$-independent coefficient from the equilibrium
calculation of (\ref{f23}), we finally  derive
eq.(\ref{e29}).
\begin{multicols}{2}

\end{multicols}

\begin{thebibliography}{99}
\bibitem{zero_freq} One needs to keep in mind that this type of
analysis is applicable only when the  $1/f$ noise is negligible.
\bibitem{cancelations}
As opposed to a single time field theory where any
contribution to an observable from  disconnected diagrams cancels with such a contribution to the partition function, in the double time formalism  each disconnected diagram vanishes separately.
This follows from the fact  that integration in the positive and the negative directions in time  gives contributions of opposite signs for disconnected diagrams.
\bibitem{pert_theory}
Such a procedure is justified if the correlation function
at hand can be correctly calculated within perturbation theory.
\bibitem{Jacob}
This is not the most general parameterization.
In our analysis  we do not include the Jacobian corresponding to this parameterization.
\bibitem{diffusion_kineq}
In order for the diffusion approximation for the kinetic equation to be
valid, the deviation from equilibrium should be small, i.e.
the energy gain between two subsequent elastic  scattering events must be small in comparison with the Fermi energy ( $(eV/\epsilon_F) (l/L) \ll 1$~).
\bibitem{cutof}  We note that in
our open boundary geometry, $E_{Th}$ appears as a natural cutoff
in our diffusons. A strong dependence on $\zeta$ will show up only
when the latter exceeds $E_{Th}$ , namely when it replaces the
Thouless energy as a cutoff.
\bibitem{wlc} Note that in the present analysis we ignore
weak localization corrections. This implies that there are no
$\langle w w \rangle$ or  $\langle w \bar{w}\rangle$
corrections to $M$, hence the current.
\end{thebibliography}
\end{document}